\RequirePackage{snapshot}
\documentclass[10pt]{article} 
\usepackage{fullpage}
\usepackage{epsfig}
\include{graphics}
\input{epsf}
\title{Fluctuations and stability in front propagation}
\author{E. Khain$^{1}$, Y. T. Lin$^{2}$,  and L. M. Sander $^{2,3}$\\
{\small$^{1}$ Physics Department, Oakland University, Rochester Mi, 48309, USA}\\              
{\small$^{2}$ Department of Physics, University of Michigan,  Ann Arbor Michigan, 48109-1040, USA}\\ 
{\small $^{3}$ Center for the Study of Complex Systems, University of Michigan, Ann Arbor Michigan, 48109-1040,  USA}
}
\begin{document}
\maketitle
\abstract{
Propagating fronts arising from bistable reaction-diffusion equations are a purely deterministic effect. Stochastic reaction-diffusion processes also show front  propagation which coincides with the deterministic effect in the limit of small fluctuations (usually, large populations). However, for larger fluctuations propagation can be affected. We give an example, based on the classic spruce-budworm model,  where the direction of wave propagation, i.e., the relative stability of two phases, can be \emph{reversed} by fluctuations.}

 \maketitle
\section{Introduction}
The study of front propagation arising from reaction-diffusion equations is a fundamental problem
in nonequilibrium physics. In this work we consider cases with two equilibria, a bistable system. If such a deterministic system forms a propagating front, we can say that the ``stable" state invades the ``metastable" state, and the theory \cite{Murray03} gives a simple criterion for which is which. 

However, we can also consider a stochastic system where fluctuations play a role, such that in the limit of small noise, we approach the deterministic system. Then the situation is not so simple, and fluctuations can play a role in determining the velocity and even the \emph{direction} of motion of the front, because fluctuations can give rise to spontaneous transitions between equilibria. That is, fluctuations can reverse the stability of the two equilibria.  In this paper we will discuss the mechanism for such reversals, and give an explicit example based on the well-known spruce budworm problem \cite{Ludwig78,Ludwig79,Murray03}.

Noise-driven transition between equilibria can be treated using the formalism of rare events. In systems with stochastic birth-death processes this approach has recently attracted considerable attention. In these systems, fluctuations lead may lead to extinction, the phenomenon that is not described by the continuum rate equations \cite{extinction}. The interplay between stochastic and deterministic effects becomes much more intriguing when spatial degrees of freedom are introduced and agents can diffuse on a lattice \cite{diffusion}. Here we focus on the role of fluctuations in the phenomenon of front propagation

\section{Deterministic definition of stability}
The standard deterministic treatment of moving fronts in reaction-diffusion systems \cite{Murray03} begins with an equation of the form:
\begin{equation}
u_t = Du_{xx} + f(u).
\label{generalform}
\end{equation}
Bistability means that 
\begin{equation}
\label{potentialV}
V(u)=-\int^u dw f(w)
\end{equation}
 has two minima, $u_1, u_3$ separated by a maximum, $u_2$. We will refer to $V$ as a potential. 

A moving front occurs when the system is divided into regions where $u$ is at different equilibria. For example, $u=u_3$ for $x<0$, $u=u_1$ for $x>0$. To analyze the dynamics, we seek a traveling front solution of Eq. (\ref{generalform}): $u(x,t) = u(\xi=x-vt)$, where $v$ is the front velocity. Substituting this into Eq. (\ref{generalform}) we get:
\begin{equation}
Du^{\prime\prime}+v\,u^{\prime}+f(u)=0,
\label{frontode}
\end{equation}
The prime indicates the derivative with respect to $\xi$. 
Multiplying Eq. (\ref{frontode}) by $u^{\prime}$ and integrating from $\xi=-\infty$ to $\infty$, we have:
\begin{equation}
v= -\int_{u_3}^{u_1} du f(u) / \int_{-\infty}^{\infty} d\xi D[u^\prime]^2 \propto V(u_1)-V(u_3).
\label{velocityeq}
\end{equation}

Note that the velocity will be positive, i.e. the state at $u_3$ will invade $u_1$ if $V_1-V_3$ is positive, so that $u_3$ corresponds to a lower potential: thus $u_3$ is stable, and $u_1$ metastable if $V_3 < V_1$. 
If  $V_1-V_3=0=\int_{u_3}^{u_1} du f(u)$ the front will not move (area rule). We refer to this as the ``stall point"; note that the stall point does not depend on $D$, but only on $V$. If $V_1-V_3$ is negative, the front will move towards negative $x$, i.e. $u_1$ will invade $u_3$. 

We will need to actually compute $v$ below, i.e. we need the denominator in the Eq. (\ref{velocityeq}).
Note that Eq.~(\ref{frontode}) can be thought of as a dynamical system; there are two stable fixed points: $u=u_1, u^{\prime}=0$ and $u=u_3, u^{\prime}=0$.  To calculate $v$ we need to find  the heteroclinic orbit connecting the two  fixed points. We do this by using a standard shooting method \cite{Sauer05}. 

\section{Stochastic definition of stability}
Equations like Eq. \ref{generalform} often arise as a mean-field description of a stochastic process such as population dynamics in spatially extended systems. To be specific, consider a birth-death process for a number, $n(j)$, of agents  that live on sites $x_j$ with birth rate $\lambda(n)$ and death rate $\mu(n)$.  Bistability means that there are three solutions to $\lambda(n)=\mu(n)$ corresponding to equilibrium population sizes on single sites, two stable and one unstable. In the spruce budworm problem, to be treated below, there are two possible stable states due to predation: $n_1$, the refuge state, and $n_3$, the outbreak state. In our example, below, we couple the sites diffusively, i.e. by introducing a rate for transfer to nearest-neighbor sites, e.g., $(n(j), n(j\pm1)) \to (n(j)-1, n(j\pm1)+1)$. 

It is well-known that the mean-field (large $n$) limit of the dynamics can be treated by scaling by some population scale, $A$, and thinking of  $n(j)/A=u(x)$ as a continuous variable. From the master equation we find that the average behavior is given by an equation of the
 form of Eq.~(\ref{generalform}) with $f = (\lambda-\mu)/A$. 
 
 However, fluctuations introduce another process, spontaneous transitions between $n_1$ and $n_3$ on a single site. The rate of transitions between the two equilibria involves a \emph{different} potential \cite{Doering05,Doering07}:
 \begin{eqnarray}
 \label{potentialPhi}
 \Phi(u) &=& - \int^u dw \ln\left(\frac{\lambda(w)}{\mu(w)}\right) \nonumber \\
 &\approx& -2 \int^u \left(\frac{f(w)}{\lambda(w)+\mu(w)}\right). 
 \end{eqnarray}
The last line is the limiting form when $f/(\lambda + \mu)$ is small (the Fokker-Planck limit). The transition time from state 3 to state 1 is of the form $t_{3 \to 1}=t_o \exp(A[\Phi(u_2) - \Phi(u_3)]$, where $t_o$ is a slowly-varying prefactor. The transition time depends on the \emph{barrier height} $\Phi_2-\Phi_3$. A similar expression holds for transitions from $u_1$ to $u_3$, $t_{1 \to 3}$. 
 
The smaller time will correspond to the smaller barrier. Based on this, we can give a different definition of stability: if $t_{3 \to 1} > t_{1 \to 3}$ we say that $u_3$ is stable, and \emph{vice-versa}. For example, $u_3$ will be more stable than $u_1$  if the barrier is larger, i.e., if  $\Phi_3 < \Phi_1$. 

This discussion is for a single site. However, spontaneous transitions can move the front if they occur at the interface between the two states. If there are transitions far from the front, and they are not too frequent, the ``hole" will quickly be filled by diffusion. We have seen examples of this in the simulations to be described below.

\section{Competition} The two potentials, $V$ and $\Phi$ are similar, but not identical. In particular, there is no reason for them to give the same result for stability. Thus a deterministic wave could move from $u_3$ to $u_1$, but noise-induced barrier climbing could cause it to move the other way. In this regime, the two effects compete. In the next section we give a numerical example where this occurs. Also, fluctuations will shift the stall point.

\begin{figure}
\centerline{\includegraphics[width=3.0in,clip=]{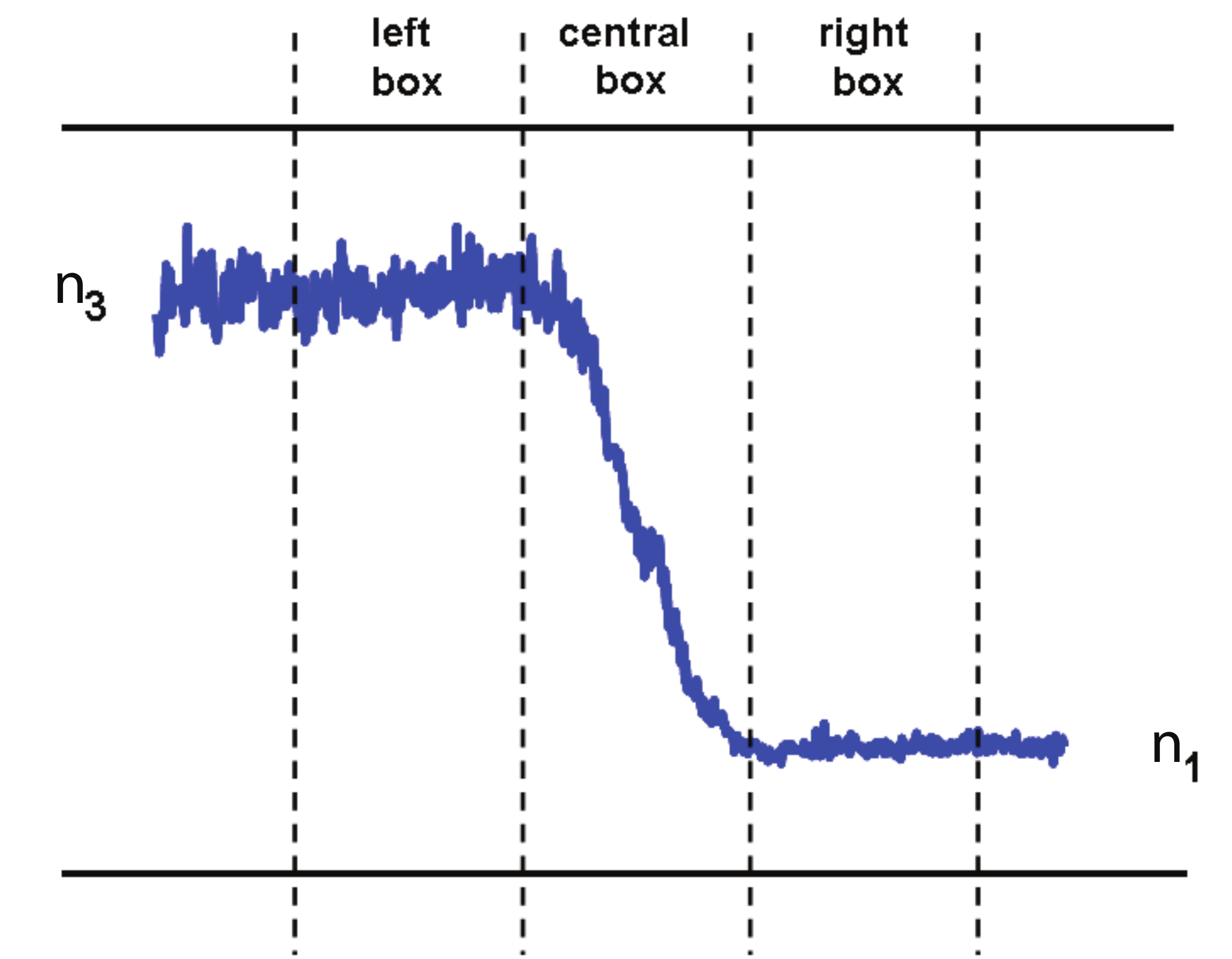}}
\caption{Front profile from simulations of the lattice model. The horizontal box size is taken to be the width of the interface, which is proportional to $\sqrt{D}$. } 
\label{front}
\end{figure}

We note that there is a natural length scale for the system given by the width of the wavefront, $w$. On dimensional grounds, $w \propto \sqrt{D}$. This leads us to imagine dividing the system into boxes of width $w$; see Figure ~\ref{front}. Focus on the central box at the front interface. Suppose the continuum front is stalled. If the probability for spontaneous jumps $u_3 \rightarrow u_1$ in the left box is larger than the probability for spontaneous jumps $u_1 \rightarrow u_3$ in the right box, the discrete front would move to the left, and the system will not be stalled. Thus the observed stall point will be different from that for a purely deterministic front. By the same token, if the two transition times are equal, the velocity will be exactly deterministic, but for other parameters the velocity will be increased or decreased by spontaneous transitions.

The relative importance of fluctuations usually depends on the size of the population. For large populations on each site, so that $n_1,n_3 \gg 1$, fluctuations are negligible. As populations decrease stochastic effects will start to affect the stall point and $v$.  Finally, for finite systems (perhaps of length $< w$) and very large fluctuations the picture of wave motion will break down altogether, and transitions of the system as a whole will be the dominant path for transitions from metastable to unstable. 

We will be interested here in the first two regimes, and the development of waves. 
All of our qualitative notions about the effects of fluctuations on waves will be illustrated by an example in the next section.

In this work we use periodic boundary conditions. We should note that with absorbing boundary conditions, $u=0$ at ends of the system, a small system may not be bistable at all \cite{Ludwig79}.

\section{Waves in the spruce budworm problem}
There is a classic model in the literature, the spruce budworm model \cite{Ludwig78,Ludwig79,Murray03}, which we will use to illustrate the effects we have discussed. The model is based on real experience in forestry where it is found that a pest which damages balsam fir trees can exist in two states, the ``refuge" state (small numbers), and the 
``outbreak" state, larger numbers. Bistability is thought to be due to the non-linear effects of birds on controlling the insect population \cite{Ludwig78}.

\begin{figure}
\centerline{\includegraphics[width=3in]{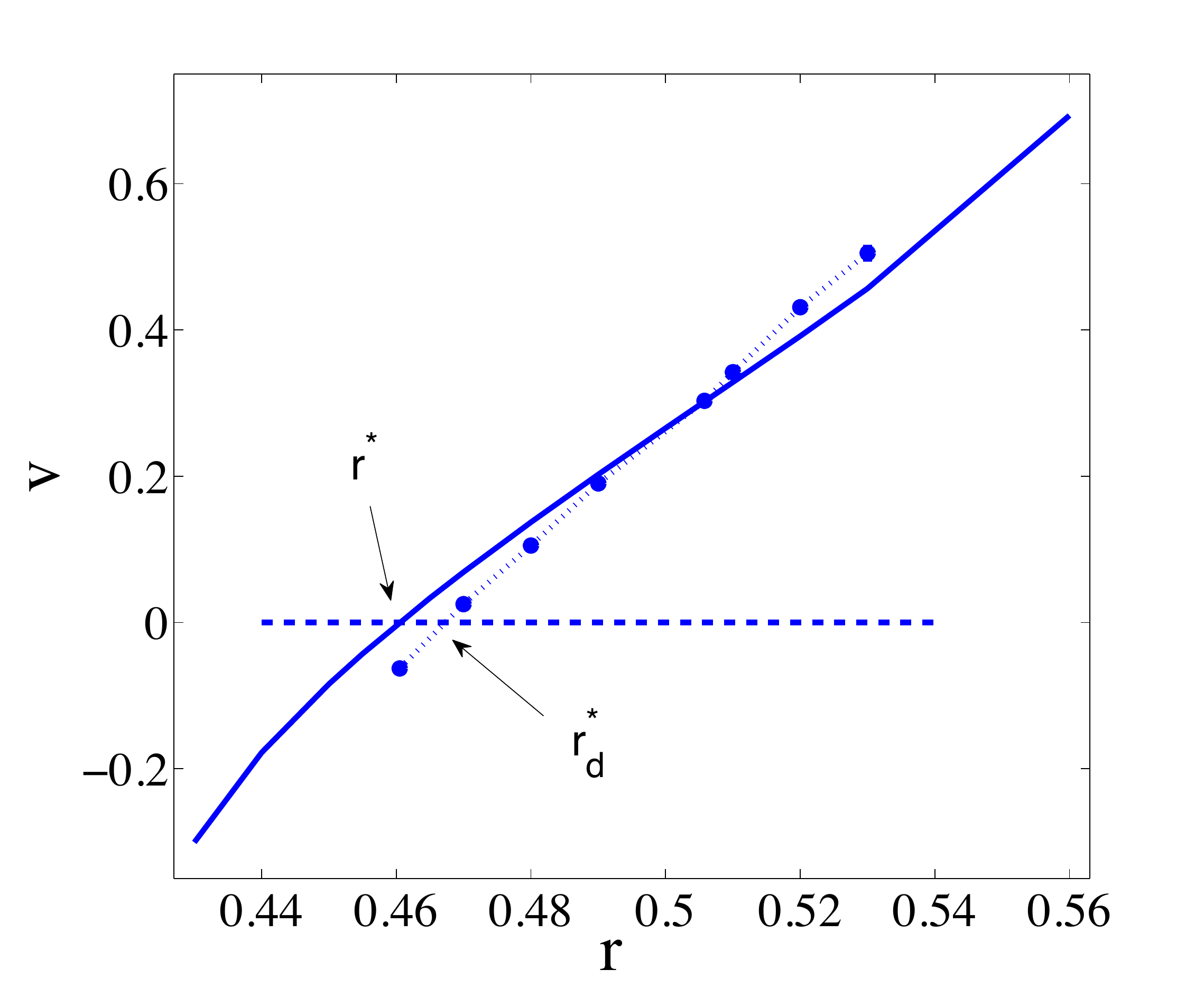}}
\caption{Velocity, $v$, as a function of birth parameter $r$  for  $q=9$, $D = 2.5$. The solid line is computed from Eq.~(\ref{budwormscaled})  and the dotted line from simulations of the discrete model. Positive velocity means that the outbreak state takes over the whole system; negative velocity means that the refuge state wins. The dashed line is $v=0$. The "continuum" stall point $r^*$ can be found from the area rule. Note that the "discrete" stall point does not coincide with the "continuum" one: $r^*_{d}>r^*$. For $r$ between $r^*$ and $r^*_d$ fluctuations reverse the velocity of the front.} 
\label{velocity}
\end{figure}

The spatially extended model \cite{Ludwig79} is written as follows:
\begin{equation}
\frac{\partial n}{\partial t} = D\,\frac{\partial^2 n}{\partial
x^2} + r_b\,n\,\left(1-\frac{n}{K_b}\right) - \frac{B\,n^2}{A^2 + n^2}.
\label{budwormspace}
\end{equation}
The first term on the right hand side of Equation 1 schematically describes migration, assuming random motion. The second term describes the usual logistic growth: insects proliferate with rate $r_b$, but there is some carrying capacity $K_b$, related to available food, which restricts exponential proliferation. The third term represents predation by birds, which saturates at large $n$: birds are not able to consume more than some maximal number of insects per unit time. When $n$ is small, the rate of predation is very small, since birds prefer other regions with a larger population of insects. Introducing dimensionless time $\bar{t}=Bt/A$, coordinate $\bar{x}=\sqrt{B/(AD)}x$,  and dimensionless number of insects $u=n/A$, we arrive at (bars are omitted for clarity)
\begin{equation}
\frac{\partial u}{\partial t} = \frac{\partial^2 u}{\partial
x^2} + r\,u\,\left(1-\frac{u}{q}\right) - \frac{u^2}{1 + u^2}.
\label{budwormscaled}
\end{equation}
Here, the proliferation parameter is $r=r_b\,A/B$ and the dimensionless carrying capacity parameter is $q=K_b/A$. In case there is no spatial dependence (a single-site problem), there exists a region in the parameter space $(r, q)$, such that there are two stable states: $u_1$ describes the normal population size and $u_3 \gg u_1$ corresponds to the outbreak \cite{Ludwig78,Ludwig79}. We consider now this region in parameter space and apply the formalism outlined above to Eq.~(\ref{budwormscaled}).  Figure ~\ref{velocity} shows the front velocity as a function of the birth parameter $r$ (solid line). For a specific birth parameter, $r^*$, the two states coexist and the front stalls, $v=0$. 

Next we analyze the same phenomena as a continuous time Markov process for agents (insects) on a lattice. Every site of the one-dimensional lattice can be occupied by any number of insects. At each time step, a site, $j$, is picked at random and then an insect on the site is picked at random. It can either jump to a neighboring site (to the right or to the left), proliferate, or die with probabilities related to the diffusion, birth and death rates on the site.
\begin{eqnarray}
 p_{birth}&=&r_b/(r_b+\mu+2D) \nonumber \\ 
 p_{death}&=&\mu/(r_b+\mu+2D) \nonumber \\
 p_{right}\quad= \quad p_{left}&=&D/(r_b+\mu+2D) \nonumber \\
 \mu&=& \frac{r_b n(j)}{K_b} + \frac{Bn(j)}{A^2+n(j)^2}. 
 \label{probabilities}
 \end{eqnarray}
The birth parameter $r_b$ is what we called $\lambda$ above. The death rate per insect, $\mu$, represents the negative terms in Eq.~(\ref{budwormspace}).  After every event, the time is advanced by $1/[n_{total}(r_b+\mu+2D)]$, where $n_{total}$ is the total number of insects. Clearly, Eq.~(\ref{budwormspace}) is the continuum analog of this discrete lattice model.

\begin{figure}
\centerline{\includegraphics[width=3.5in,clip=]{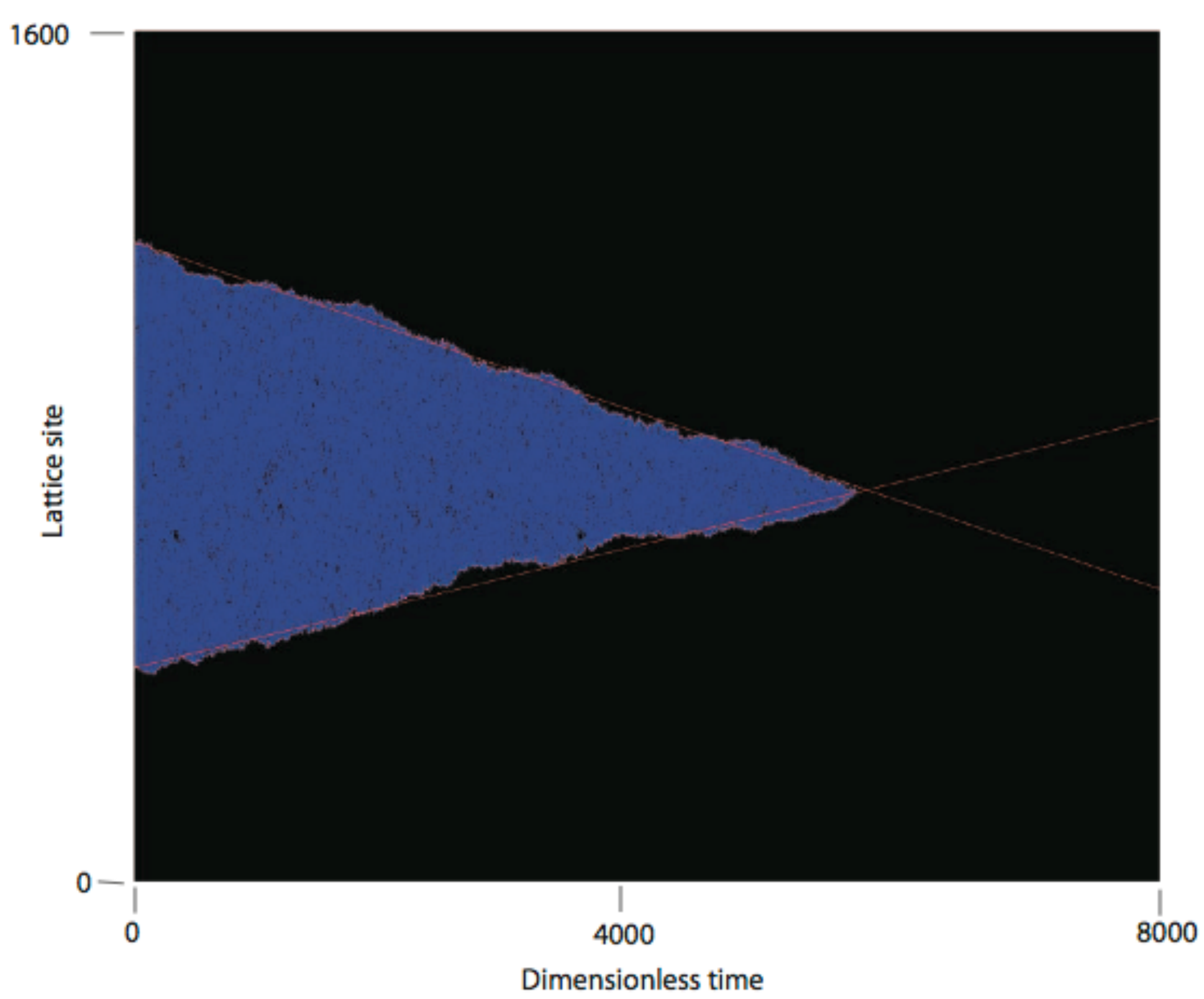}}
\caption{Space-time diagram for the single simulation. Each point on the diagram shows the number of particles $n$ in the specific site at some specific time; the color is related to value of $n$. Initially, a  region of $n_3$, the outbreak state, was surrounded by the $n_1$ state. After the two fronts propagated toward each other, the refuge state took over the whole system. The velocity is computed from the slope of the front interface. The parameters are $q=9, D=2.5, r=0.4605$.} \label{spacetime}
\end{figure}

We performed many simulations of the discrete model. To compute the front velocity, we first did a time-average for a single run; this was done by plotting the space-time diagram, see Figure~\ref{spacetime}, and computing the slope. Then we averaged the results over many ($50-100$) simulations. Figure ~\ref{velocity} presents the front velocity as a function of the birth parameter $r$ (dotted line). Strikingly, the discrete stall point $r^*_{d}$ is not equal to the continuum stall point: $r^*_{d}>r^*$. 

This is precisely the result presented above: since  the jump probability from $n_1$ to $n_3$ is not equal to the jump probability from $n_3$ to $n_1$, spontaneous jumps contribute to the velocity. Since the probability decreases exponentially with the population scale $A$, the stall point shift $r^*_{d}-r^*$ should tend to zero when the number of particles increases. Simulations of the discrete model show that this is indeed the case: $r^*_{d}$ tends to its continuum value $r^*$ as $n$ increases.

\begin{figure}
\centerline{\includegraphics[width=3.0in,clip=]{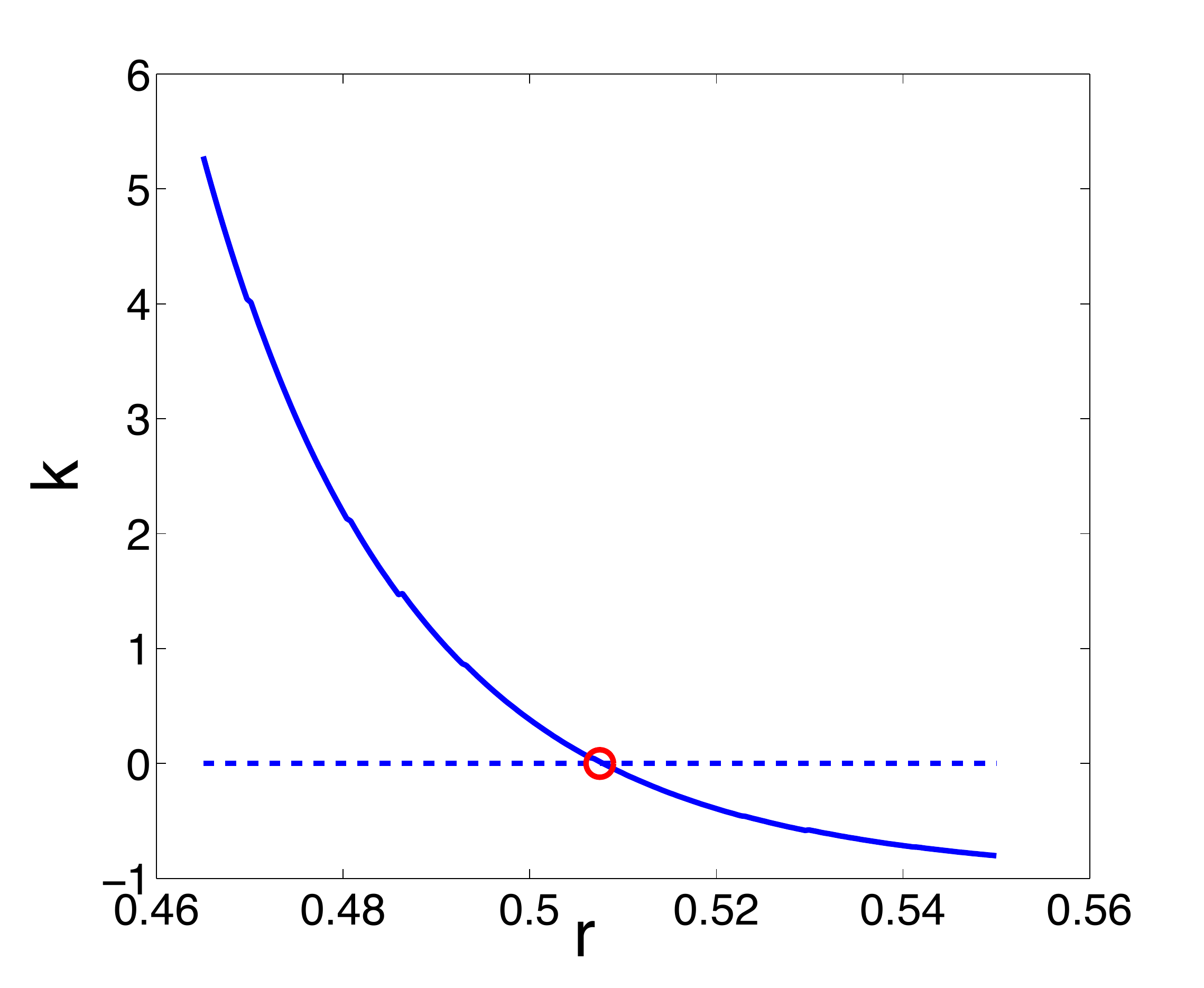}}
\caption{The parameter $k=(t_{1\to 3}/t_{3 \to 1}-1)$ as a function of $r$. The circle denotes the value of $r$ ($\bar{r}\simeq 0.5075$) when the two times are equal. The other parameters are $q=9$, $D=2.5$, $A=7$.} \label{transition}
\end{figure}

We suggested above that the "continuum" and "discrete" velocities in Figure ~\ref{velocity} should be equal when the mean transition time from $n_1$ to $n_3$, $t_{1 \to 3}$ equals the time for the reverse transition, $t_{3 \to 1}$. Figure ~\ref{transition} shows $k=[t_{1\to 3}/t_{3 \to 1}-1]$ versus the birth parameter $r$. Note that the times are equal at approximately $\bar{r}=0.5075$. This suggests that at this value of birth parameter $r$, there is no stochastic correction to the continuum velocity. Our simulations support this prediction: Figure ~\ref{velocity} shows that the continuum and the discrete velocities are equal at $r=\bar{r}$.

\begin{figure}
\centerline{\includegraphics[width=3.0in,clip=]{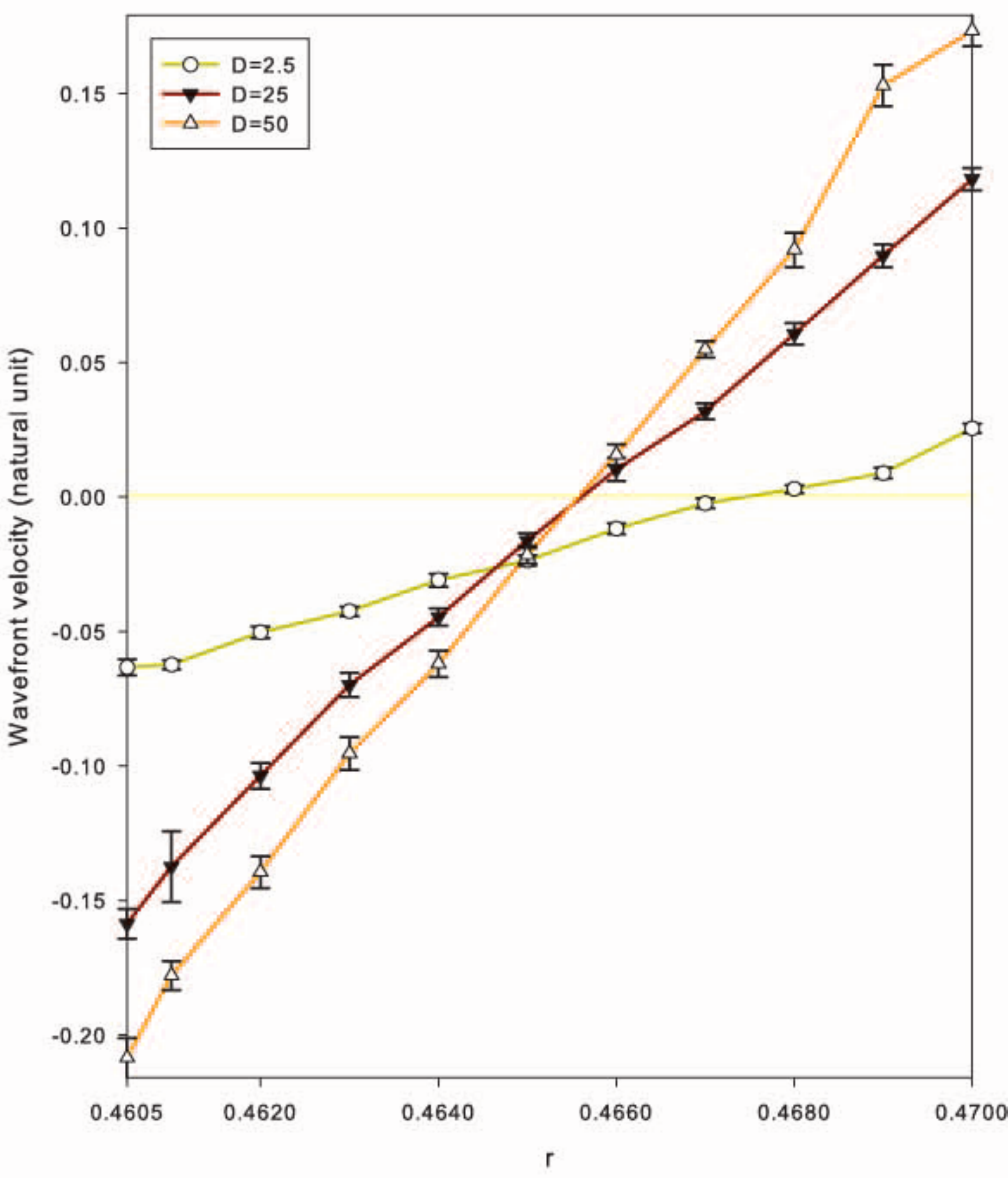}}
\caption{Velocity of front propagation from the  discrete lattice model, for various values of $D$ for $q=9, A=7$.} \label{velocity2}
\end{figure}

As was mentioned above, the stall point obtained from the continuum theory, $r^*$, does not depend on the diffusion coefficient. This is not the case in discrete lattice system: Figure ~\ref{velocity2} shows the $v(r)$ dependence for various diffusion coefficients; $r^*_{d}$ depends on $D$. We might expect that  the shift $r^*_{d}-r^*$  would tend to zero as the diffusion coefficient increases, and effectively coarse-grains the system. We might argue as follows: stochastic transitions contribute to the front velocity  when they occur in all the sites inside the left or the right box, see Figure ~\ref{front}. The typical front width scales as $\sqrt{D}$. The larger the box, the smaller the probability for such a collective jump. Therefore, as $D$ increases, the stall point shift  should tend to zero. However, Figure \ref{velocity2} shows a small discrepancy for large $D$ and no sign of convergence to $r^* =0.4605$. In fact, our simulations (not shown) reveal that the equilibria, $n_1, n_3$ for the discrete model do not approach those for the continuum model for large $D$, but are uniformly shifted by a small amount. We do not understand the large $D$ limit for this system. 

There is another effect that we observed in our simulations. We are dealing with quite small numbers, $n$, so that spontaneous transitions are reasonably common.  In this case ``islands" of the stable state can appear ahead of the front and either disappear, presumably because they are smaller than some critical nucleation size, or be enveloped by the advancing front. This has been observed previously  in other stochastic wave-front  problems for similar reasons \cite{Islands}.  

\section{Discussion} It is well known that fluctuations can strongly affect front propagation for situations when a stable state invades an unstable state \cite{Brunet97,Kessler98,Panja04}, the ``pulled case". Here we are dealing with ``pushed" fronts, and the qualitative effects are more subtle, and only occur in a limited region of parameter space. Nevertheless, we have shown that they can occur. They should be considered whenever small populations of discrete agents are involved in a spatially spreading process.

\section{Acknowledgments}
E.K. thanks B. Meerson and Y. Louzoun and LMS thanks C. Doering and D. Lubensky for useful discussions.
\bibliographystyle{eplbib}

\end{document}